\documentclass[12pt]{iopart}
\usepackage{graphicx}
\usepackage{dcolumn}
\usepackage{epstopdf}

\begin{document}

\title{Disorder-driven loss of phase coherence in a quasi-2D cold atom system}


\author{M.C. Beeler$^1$, M.E.W. Reed$^1$, T. Hong$^1$\footnote{Present Address:Shanghai Advanced Research Institute, Chinese Academy of Sciences, Shanghai, 201203, China} and S.L. Rolston$^1$ }

\address{Joint Quantum Institute, University of Maryland and National Institute of Standards and Technology, College Park, MD 20742, USA}

\ead{rolston@mail.umd.edu}

\begin{abstract}

We study the order parameter of a quasi-2D gas of ultracold atoms trapped in an optical potential in the presence of controllable disorder. Our results show that disorder drives phase fluctuations without significantly affecting the amplitude of the quasi-condensate order parameter. This is evidence that disorder can drive phase fluctuations in 2D systems, relevant to the phase-fluctuation mechanism for the superconductor-to-insulator phase transition (SIT) in disordered 2D superconductors.
\end{abstract}

\pacs{05.30.Jp, 67.85.-d, 67.85.Hj}

\maketitle

\section{Introduction}
Although it has been 25 years since the discovery of high-temperature (high-T$_c$) superconductivity in cuprates, the nature of the superconducting phase transition surprisingly remains an open question. This is in part due to the effects of disorder in two-dimensional (2D) systems: 2D physics is well understood in the context of Berezinskii-Kosterlitz-Thouless (BKT) theory \cite{Thouless1973, Berezinskii1970}, but the addition of disorder can make even the phase diagram uncertain \cite{Phillips2003}.  Disorder-driven phase fluctuations in 2D systems may produce a Bosonic-type phase transition between a superconducting and insulating state in superconducting thin films \cite{Girvin1990,fisher1990, Markovic1998}; there is mounting evidence that many phase transitions in high-T$_c$ superconductors are indeed Bosonic \cite{Bozovic2011,Ong2010} in nature. As it is difficult to access the phase of the order parameter in these systems, other mechanisms behind the transition (e.g. pair-breaking) are difficult to rule out. This theoretically challenging problem is well suited for quantum simulation. We replace the high-T$_c$ superconducting system with a well-controlled ultracold Bosonic system, where we gain access to the phase of the order parameter through matter-wave interference.

Two-dimensional systems behave very differently from 3D systems. For instance, in 2D, there can be no long- range order. However, this does not preclude the existence of a superfluid state for low temperatures, with this superfluid destroyed at a critical temperature T$_{BKT}$. At non-zero low temperatures, thermal excitations exist only in the phase of the order parameter of the superfluid, taking the form of long-wavelength phonons and vortices - quantized circular superfluid  flow around a small spot of non-superfluid. Below T$_{BKT}$, vortices are bound into pairs with opposite direction of circulation, and the transition to a normal fluid (or gas) occurs after these vortex pairs unbind. These phase fluctuations are believed to be the mechanism for the BKT transition observed in 2D ultracold atom systems \cite{Dalibard2006, Phillips2009}, as well as the temperature-driven SIT in superconducting thin films \cite{Girvin1990,fisher1990}, and may be relevant to the zero-temperature superconducting doping phase transition in high-T$_c$ cuprate superconductors \cite{Bozovic2011}. In this work, we use laser speckle to controllably add disorder to a quasi-2D Boson system of Rb atoms trapped in two planes of an optical lattice. By analyzing fringes due to matter-wave interference of the released atoms,  we show that disorder predominantly drives phase fluctuations, but not fluctuations in the amplitude of the quasi-condensate order parameter, expected to be partially superfluid. This work complements other studies of disorder in ultracold atom systems \cite{Inguscio2008} in 1D \cite{Hulet2010, Aspect2008}, 2D \cite{Bouyer2010}, and 3D \cite{DeMarco2009}.

\section{Experiment}
\begin{figure}
\begin{center}
\includegraphics[scale=1.6]{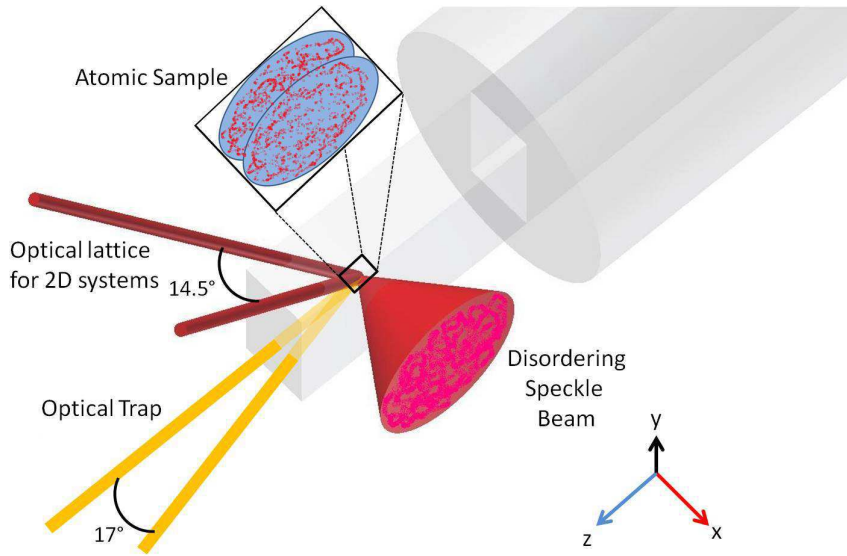}
\caption{{\bf Overview of the experimental setup.} The atoms are confined in a combination of optical trap, shallow-angle optical lattice, and disordering beam. Gravity is in the $-y$ direction, while the two planar systems are perpendicular to the $x$-direction. The camera used to take images is situated above the sample. \label{fig:setup}}
\end{center}
\end{figure}

Our experiment begins with a 3D Bose-Einstein condensate (BEC) of $^{87}$Rb atoms confined in a crossed-beam optical dipole trap of frequencies $(\omega_x, \omega_y, \omega_z)/2\pi \approx (125, 125, 30)$ Hz.  A shallow-angle optical lattice divides this 3D BEC into two nearly identical quasi-2D ultracold atom samples confined at two lattice nodes \cite{Dalibard2006} separated by 3.1 $\mu$m, while an additional beam introduces a disordering potential (Fig. \ref{fig:setup}). We adjust the temperature of the two samples between 25 and 400 nK by changing the final optical trap depth, which simultaneously tunes the total atom number from 15 X $10^4$ to 80 X $10^4$. The lattice provides a confinement frequency in each of the planes of $\omega_x/2\pi$ = 7.64(4) kHz, which is greater than our largest chemical potentials $\mu/h \sim 3$  kHz ($h$ Planck's constant) and similar to $k_B T/h$ ($k_B$ Boltzmann's constant) at our highest investigated temperatures $T$, indicating that each plane is in the quasi-2D limit. The dimensionless 2D interaction constant $\tilde{g}\approx0.21$ \cite{Dalibard2009}. Tunneling between  planes is negligible on the timescale of the experiment, and small occupations in lattice nodes outside the central pair are insignificant.

The disorder in the potential experienced by the atoms is directly proportional to the intensity in the disordering beam \cite{Inguscio2008}. In order to create spatially small intensity features across our $\sim$20x90 $\mu$m planes (estimated Thomas-Fermi diameter of the planes using the perpendicular lattice frequency  \cite{SmithBook}), we pass a 777.3 nm laser beam through a phase diffuser and focus the beam to a Gaussian waist of 90 $\mu$m on the atoms using a high numerical aperture lens (NA = 0.45). The measured Gaussian waists of the autocorrelation function of our disorder \cite{Bouyer2006} are 3.2 $\mu$m along the beam axis and 530 nm transverse to the axis, providing an accurate measure of the (randomly varying) average feature size in each direction (see Supplementary Material). Since the disorder-inducing beam is perpendicular to the 2D planes, the feature size across the planes is 530 nm. Additionally, as the spacing between the planes is on the order of the correlation length along the beam, we expect that each plane experiences approximately the same disordered potential. 

The 777.3 nm lattice light is turned on with two linear ramps of 500 ms and 100 ms to maintain adiabaticity, followed by a 200 ms hold to ensure equilibrium. During the first (second) linear ramp, the lattice attains a final potential depth of $U_0/h$ = 6.1$\pm$0.3 (290$\pm$20) kHz. The disorder strength is changed linearly from 0 to its final value $\Delta_f/h$ between 0 and 4.2$\pm$0.4 kHz (determined by the final power in the disordering beam) during the second ramp. After the hold time, we shut off all of the beams in $<$200 $\mu$s. We measure the expanded density distribution of the planes with absorption imaging along the $y$-direction  after 22 ms of expansion and extract information about the distribution and interference contrast.

In the absence of disorder, before shutting off the potentials, the density distribution smoothly decreases from a maximum at the center of the cloud to zero at the edge. Using a local density approximation (LDA), the local properties of the system are a function only of the position-dependent 2D phase space density $D(r) \equiv n(r)\lambda_{dB}^2(T)$, with $n$ the density and $\lambda_{dB}$ the thermal deBroglie wavelength \cite{Dalibard2009, Chin2011}. At the edges of the cloud, where $D$ is low, the gas is thermal. As $D$ increases towards the center of the sample, the density fluctuations are suppressed, leading to a pre-superfluid phase which supports vortices and can be considered a quasi-condensate - a BEC with a fluctuating phase \cite{Cornell2010, Phillips2009, Svistunov2000, Blakie2009}. Finally, when $D>D_c$ (equivalent to T $<$ T$_{BKT}$ at the local density), the system undergoes the BKT phase transition to a superfluid as thermally activated vortices pair (as observed in \cite{Dalibard2006}).

\begin{figure}
\begin{center}
\includegraphics[scale=1.5]{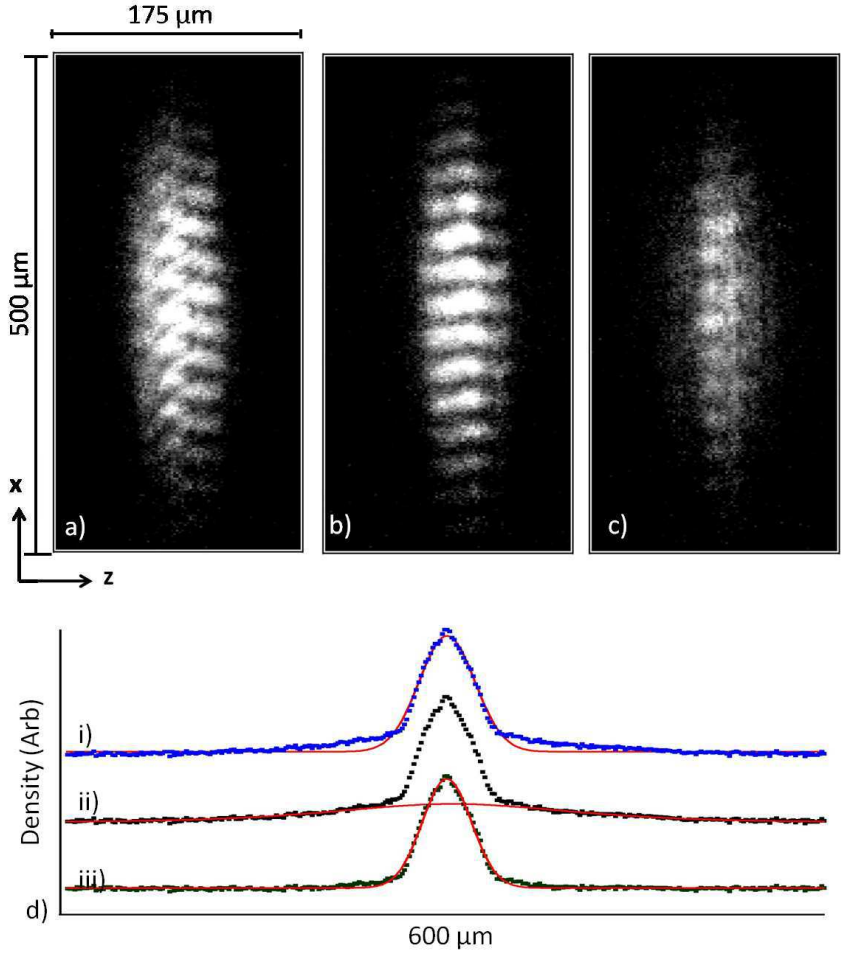}
\caption{{\bf Sample absorption images.} Images after 22 ms expansion show the effect of a vortex on the fringe pattern (a), straight fringes (b), and the effect of $\Delta/h = 7$ kHz strength disorder (c). d) shows a cut along $z$ (integrated along $x$) with a bi-modal distribution fit by a single Gaussian (i), with a Bose function fit to the thermal portion (ii), and with a Gaussian fit to the central peak with the thermal portion removed (iii). A well-separated vortex originally present in one of the planes is indicated by the dislocation of the phase of the fringe pattern near the center of the cloud in (a). \label{fig:absorption}}
\end{center}
\end{figure}

When the potentials are shut off, the clouds expand very quickly along the direction of tight confinement ($x$), producing a nearly interaction-free expansion \cite{Dalibard2011, Cornell2010, Bourdel2011}. In the $z$-direction, we see a bimodal distribution (Fig. \ref{fig:absorption}b, d). Atoms in the low-density region of the cloud  expand ballistically according to the temperature of the sample, producing a broad Gaussian background. Atoms in higher density regions near the center of the cloud will expand with a characteristic momentum set by the inverse position-space coherence, giving a sharp peak in the expansion along $z$ \cite{Dalibard2009, Cornell2010}. All of our data has $D(0) > D_c \approx 7.4$ \cite{Svistunov2001, Dalibard2009}, so there are atoms in the central peak in all images. Along $x$, the two planes completely overlap after expansion and display an interference pattern \cite{Dalibard2006, Ketterle1997} (Fig. \ref{fig:absorption}). If all of the atoms were 100$\%$ Bose-condensed, then the contrast of the interference fringes would be 100$\%$ \cite{Demler2006}. Thermal phase fluctuations prevent BEC formation and decrease this visibility through both phonons and vortices, even after subtracting off the broad background of non-coherent atoms. The properties of the non-disordered trapped quasi-2D Bose gas and its expansion properties have been extensively studied \cite{Dalibard2006, Dalibard2011, Cornell2010, Chin2011, Dalibard2009, Bourdel2011}. 

We now discuss the effect of disorder on the properties of a quasi-2D Bose gas. The presence of disorder can modify both the  properties of the trapped gas and the properties of the expansion. The density distribution in the trap should take the shape of the disorder (more atoms settle into the wells of the disorder), although this effect is expected to be screened by repulsive interactions between the atoms \cite{Gunn1990}. We expect that the disorder will also disrupt the phase coherence of the trapped sample. We would like to get information about the properties (particularly phase coherence) of the trapped atoms by looking at the atoms after 22 ms of expansion.

To disentangle the effects of density and phase modifications in the trapped gas and their effects on our measurements, we have first verified that the disorder has no effect on the temperature (see Supplementary Material). If only the phase coherence of the trapped samples becomes disordered, we expect that the integrated visibility of the interference pattern observed after expansion will decrease. A smaller local trapped density (induced by a local minimum in the disorder) will also decrease the $local$ visibility after expansion, but if those missing atoms remain in the central peak during the expansion, the $integrated$ visibility of the central peak will not change. The disorder could also disrupt overall coherence such that atoms which would have appeared in the central peak after expansion are redistributed to the wings of the distribution during the expansion. We can measure this effect by counting the fraction of atoms in the central peak in our images.  

\section{Analysis}
\begin{figure}
\begin{center}
\includegraphics[scale=0.75]{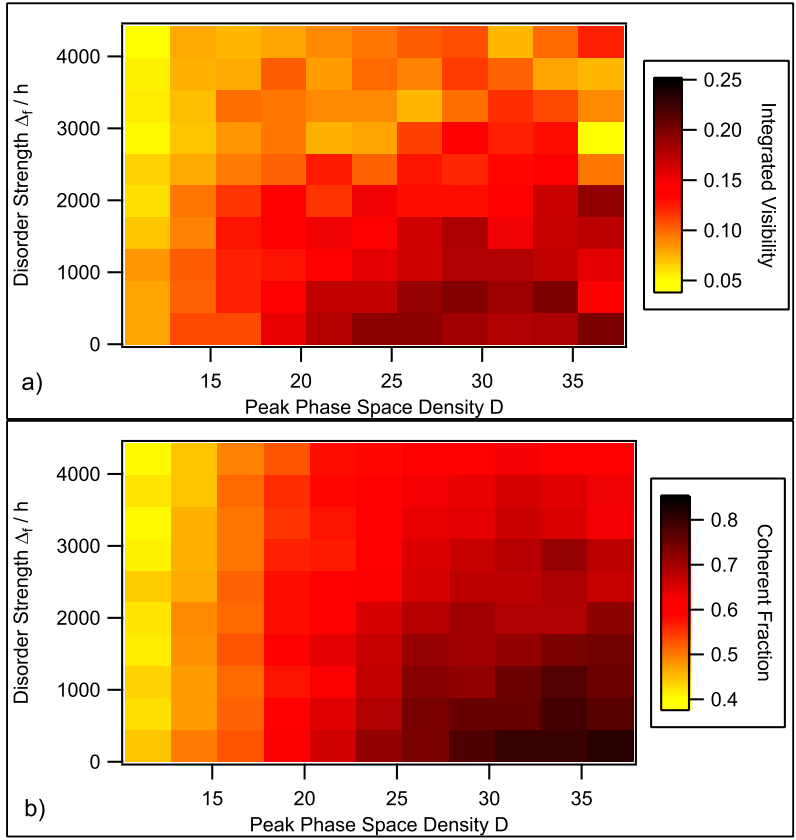}
\caption{{\bf Disorder-Phase Space Phase Diagrams} Color maps showing integrated visibility over the central portion of the interference pattern (a) and coherent fraction (b) in our images as a function of disorder strength and phase space density. The color maps are qualitatively similar but the drop in coherent fraction at high disorder and high $D$ (top right corner of the color maps) is much less than the corresponding drop in visibility. \label{fig:phasediagram}}
\end{center}
\end{figure}

Each image of the expanded cloud is analyzed to obtain coherent fraction, temperature, total atom number, and integrated fringe visibility. Total atom number is obtained by integrating the density distribution in each image. To obtain temperature and coherent atom number, we use the observed two-component distribution along the $z$-axis in the images. We integrate along the $x$-direction and then exclude the central portion of the cloud, fitting the remaining distribution along $z$ to a Bose function. The width of this Bose function is used to obtain the temperature. We subtract off the Bose fit, assuming that the thermal component in the $x$-direction has the same width as the coherent component, and then integrate the remaining atom density to obtain the coherent atom number. The integrated fringe visibility is determined by taking the 1D Fourier transform of the coherent atoms along $x$ and finding the amplitude $A(z)$ and phase $\phi(z)$ of the Fourier peak at the frequency of the fringes.

Figure \ref{fig:phasediagram}a) shows the average fringe visibility $\int{A(z)e^{i \phi(z)}dz}$ integrated over 120$\%$ of the Gaussian width of the central portion of each cloud at different peak phase space densities and disorder strengths. The $D$ shown here is what the peak $D$ in each of our planes would be at the measured temperature and atom number in the absence of disorder, calculated with a local density approximation (LDA, see supplementary material). Since the local properties of the (non-disordered) gas are completely determined by $D$, this is the appropriate quantity to use along one axis. We have also measured the fraction of atoms in the coherent central part of the measured bi-modal distribution (Fig. \ref{fig:phasediagram}b). We observe a similar phase diagram for this central coherent fraction as for the integrated visibility. However, the effect of the disorder is not as strong, most notably in its effect at high $D$ and high $\Delta_f/h$. 

\begin{figure}
\begin{center}
\includegraphics[scale=0.75]{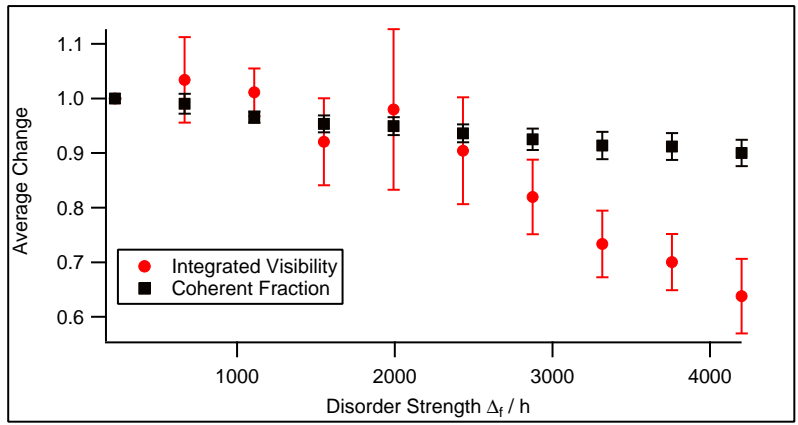}
\end{center}
\caption{{\bf Fractional Change Due to Disorder} Average fractional change for the coherent fraction and integrated visibility as a function of disorder strength. Each data point is the average over $D$ at that disorder strength, and error bars are 90$\%$ confidence intervals for that average. \label{fig:visbec}}
\end{figure}

We can quantify the difference in fractional decrease in visibility vs. fractional decrease in coherent fraction as a function of disorder strength (Fig. \ref{fig:visbec}). Ideally, we would like to compare the asymptotic values that the visibility and coherent fraction approach at high $D$ and plot them as a function of disorder. However, it is unclear if either visibility or coherent fraction is approaching a constant value at high $D$, even in the absence of disorder. Instead, we average the fractional change across all phase space densities at each disorder strength. This serves to weaken the signal, as we expect the role of disorder to decrease at low $D$. Still, we find that once the disorder strength exceeds 4 kHz, the fractional decrease in visibility is $\sim$30$\%$ more than the fractional decrease in coherent fraction.

In regions of the sample where $D$ is high enough to be at least quasi-condensed, we can describe the system with an order parameter having both an amplitude and a phase - these are the atoms in the central peak of the expanded distribution. An explanation of Fig. \ref{fig:visbec} is that the disorder has affected the phase of this order parameter in each system while not affecting the amplitude of the order parameter nearly as much, as predicted by the dirty Boson model \cite{Markovic1998, fisher1990, Girvin1990}. This theory shows that disorder generates phase fluctuations in a low temperature superfluid that drive the system into an insulating state. In a quasi-2D ultracold gas, it is expected (but not yet experimentally verified) that the gas is locally superfluid when $D > D_c$, which may make this model applicable. Until now, the effect of disorder on quasi-2D superfluids described by this model has been limited to liquid $^4$He thin films \cite{ Hallock2006} or thin film superconductors \cite{Goldman1998}, with no access to the phase of the superfluid order parameter. In our system, we can measure the phase of the order parameter, but we do not observe a clear indication of a transition at high disorder strengths. Instead, we see the visibility level off to a $D$-dependent value at high disorder strength. 

This leveling may be related to confinement or finite-sized system effects.  The relevant length scales are the system size (20 x 90 $\mu$m), the disorder length scale ($\sim$0.53 $\mu$m), and the healing length (0.25 $\mu$m). We can investigate the effect of larger disorder correlation lengths by decreasing the numerical aperture of our disorder optical system. Models based on random Josephson junctions \cite{Garland1991} or percolation \cite{Essam1980} may also help explain these results, but a clear theoretical understanding is made difficult by the non-uniform system density and the exact nature of the coherence in the central peak. Other ultracold atom disorder experiments may have also observed similar decoupling of phase and condensate fraction \cite{Demarco2010, Inguscio2010, Inguscio2008}, but there are significant differences in the dimensionality and underlying structure in these experiments, making direct comparisons difficult.

We observe well-separated vortices in some of our images. The statistics of these vortices may be used to further characterize the effect of disorder in 2D systems, possibly indicating pinned vortices, as in type II superconductors \cite{Vinokur1994}. Preliminary measurements suggest that the percentage of realizations with one or more vortices present does not depend on the amount of disorder, but more data must be taken to support any conclusions. We can also translate the disorder relative to the confining potential, allowing for the exploration of vortex pinning and transport. We have shown that disorder can drive phase fluctuations in quasi-2D systems, a direct observation of the mechanism predicted for the disorder-driven SIT in the dirty Boson model. This has relevance to systems such as superconducting thin films and high-T$_c$ cuprate superconductors, which may undergo phase transitions by this mechanism. This result may bring us closer to understanding the mechanism behind phase transitions in high-T$_c$ superconductors.

\ack{ We acknowledge E. Edwards for help with the initial implementation of the experiment. The program was funded with support from the JQI, ARO, NSF, and PFC at JQI. M.C.B. acknowledges support from NIST-ARRA. }

\appendix
\section*{Appendix}
\setcounter{section}{1}
\subsection{Disorder Calibration.}
The correlation length of the disorder was calibrated by imaging the disordered intensity pattern $I(\bf{r})$ using a microscope objective and CCD camera. We computed the autocorrelation function $C(\mathbf{r}) = \left<I(\mathbf{r_0})I(\mathbf{r_0+r})\right>$, where the brackets represent an average over $\bf{r_0}$, and used the Gaussina width of the function to determine the correlation lengths. For the transverse autocorrelation length,  we used a single image of the beam at the focus of the final lens. To obtain the axial autocorrelation length, the imaging system was mounted on a translation stage with $\sim$30 nm step size, and a series of images was taken and then corrected for transverse jitter before the autocorrelation function of a slice along the axis was computed.

To calibrate the strength of the disorder, we pulsed the disordered potential on a thermal cloud of atoms and observed the increase in momentum distribution width. If the pulse is short compared to atomic motion timescales, it gives an impulse $\Delta \vec{P}(\mathbf{r}) = \vec{\nabla} U(\mathbf{r})$, where $U(\mathbf{r})$ is the potential strength of the disorder, proportional to the intensity of the beam. We modeled the process using a Monte Carlo simulation and compared the actual increase in momentum distribution width to our simulation in order to obtain disorder strength as a function of power in the disordering beam.

\subsection{Color Maps}

The bin sizes for the color maps were chosen so that almost all bins had 2 or more points, with some bins in the middle of the maps containing as many as 50 points. A total of 1300 images were analyzed to produce the color maps.  Each bin is colored according to a Gaussian weighted average over the entire data set, with the Gaussian width set to 1/3 of the bin size.

\subsection{Disorder Effect on Temperature Measurement}
\begin{figure}
\begin{center}
\includegraphics[scale=0.75]{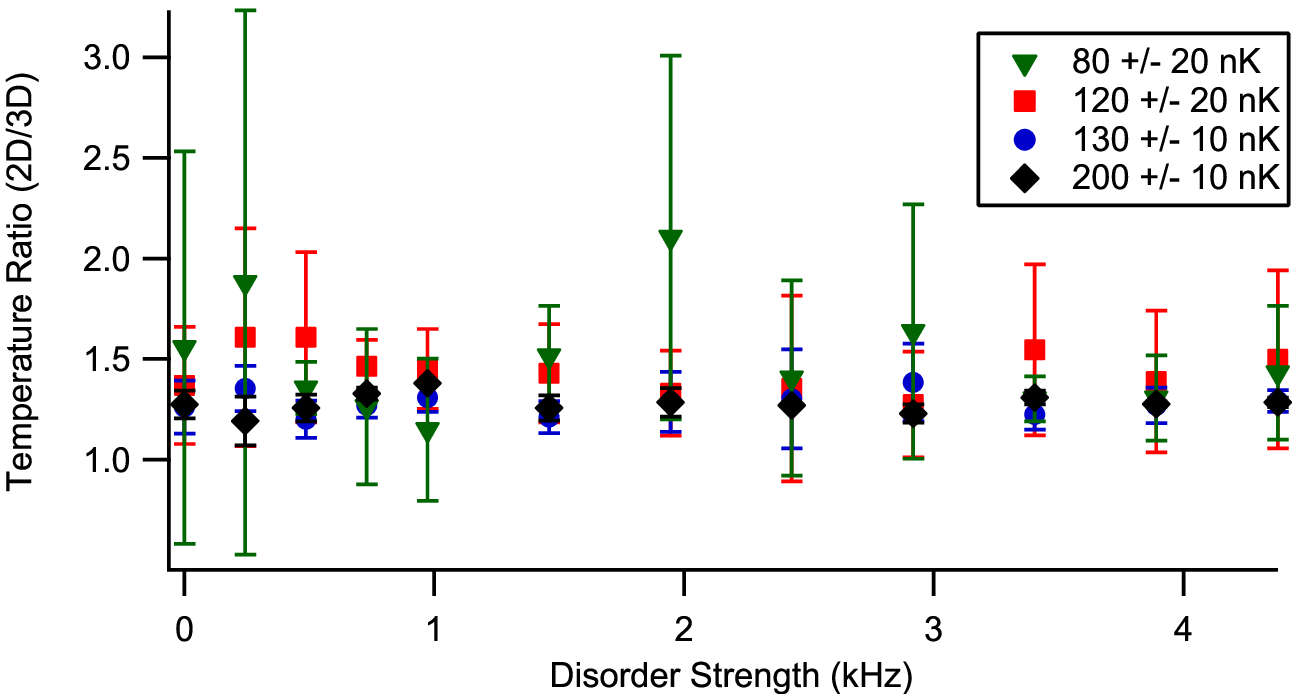}
\end{center}
\caption{{\bf Effect of Lattice and Disorder on Temperature} Ratio of temperature measurements in concurrent shots with and without the lattice and disorder. Each point is the average of $\sim$ 5 shots at the same value, with error bars the standard deviation. For a given temperature set, the quoted temperature is the average temperature of the shots without the lattice and disorder, with the standard deviation giving the error. The flatness of the curves indicates that increasing disorder has no effect on the measured temperature of the sample. \label{fig:temp}}
\end{figure}

We took several hundred alternating images with our lattice and disorder to determine the effect of the lattice and disorder on the measured temperature of the thermal cloud. The first image was taken with our typical procedure described above, while the second image in each set followed the same timing, but the lattice and disordering beams were never turned on. For the 3D BEC image, the center was masked and we fit 1D cuts of the thermal background in two directions to Gaussians to extract the temperature. We show in Fig. \ref{fig:temp} the ratio of the temperature measured in concurrent shots for a variety of disorder strengths. The ratio looks flat as a function of disorder, indicating that higher disorder strengths have a negligible effect on our temperature measurement. We repeated the measurement at various average temperatures to be sure that there was no temperature dependence to the curve. A combination of adiabatic heating and spontaneous emission heating increase the temperature in the lattice and disorder by a factor of $\approx$ 1.4 from its 3D BEC value.

\subsection{Phase Space Density and Chemical Potential}
Once each of our images is analyzed to obtain temperature and total atom number, these quantities (along with the properties of the trapping potential) can be used to calculate phase space density and chemical potential in the analogous disorder-free system. It is easiest to divide quasi-2D systems into three distinct phases: normal, fluctuation, and Thomas-Fermi regimes, corresponding to low, medium, and high densities. In the normal regime, the density can be obtained using the following procedure from Refs. \cite{Cornell2010, Dalibard2011}. 

The 2D in-plane density $n(\vec{r})$ in each vibrational level $j$ (assumed to be quantum harmonic oscillator eigenstates) along $x$  is
\begin{equation}
\label{eq:density}
n_j(\vec{r}) = -\mathrm{ln}\left(1-e^{\mu_j(\vec{r})/k_BT}\right)/\lambda_{dB}^2,
\end{equation}
with
\begin{equation}
\label{eq:chempot}
\mu_j(\vec{r})=\mu^{global}-\frac{1}{2}m(\omega_y^2y^2+\omega_z^2z^2)-j\hbar\omega_x-\displaystyle\sum\limits_{l}2\left(\frac{4\pi\hbar^2}{m}a_sf_{jl}n_l(\vec{r})\right),
\end{equation}
where $k_B$ is Boltzmann's constant, $\lambda_{dB} = \sqrt{2\pi\hbar^2/(mk_BT)}$ is the thermal deBroglie wavelength, $\mu^{global}$ is the peak chemical potential, $m$ is the mass of $^{87}$Rb, $\omega_i$ is the harmonic trapping frequency in the given direction, $\hbar$ is Planck's constant, and $a_s$ is the s-wave scattering length of $^{87}$Rb. In Eq. \ref{eq:chempot}, $f_{jl}$ is the normalized density overlap integral between harmonic oscillator eigenstates $j$ and $l$. Eq. \ref{eq:density} can be solved via an iterated fixed point method, and converges after a few iterations. Note that phase space density $D_j(\vec{r}) = n_j(\vec{r})\lambda_{dB}^2$.

In the high-density Thomas-Fermi region of the cloud, the density is given by Eq. \ref{eq:density}, except that the factor of $2$ in front of the sum in Eq. \ref{eq:chempot} is not there. That factor of $2$ is there in the normal phase because density fluctuations are important, so that $\left<n^2\right>=2\left<n\right>^2$. When the density gets very high, density fluctuations are suppressed, and $\left<n^2\right>=\left<n\right>^2$, removing the factor of 2 in the sum term of Eq. \ref{eq:chempot}. 

In between these two densities, the relationship is more complicated. In the strictly 2D case, where only the lowest harmonic oscillator eigenstate is occupied, the phase space density in this fluctuation regime is calculated with a Monte Carlo method, and the results are tabulated by Prokof'ev and Svistunov \cite{Svistunov2002}. These results were extended to the quasi-2D case for finite occupation of excited vibrational modes by Holzmann, Chevallier, and Krauth \cite{Krauth2010}. They found that one could define a correlation density $\Delta D$ - the difference between the true phase space density in the fluctuation regime and what is predicted by Eq. \ref{eq:density}. The correlation density is a function only of the mean field gap
\begin{equation}
\Delta_{mf}=-\mathrm{ln}\left(1-e^{-D_0}\right), 
\end{equation}
with $D_0$ the ground vibrational state phase density calculated using Eq. \ref{eq:density}, and is given with 10$\%$ accuracy up to the BKT transition density $D_c$ to be
\begin{equation}
\Delta D\simeq\frac{1}{5}\left(-1+\frac{\tilde{g}}{\Delta_{mf}}\right)\frac{1}{1+\pi\Delta_{mf}^2/\tilde{g}^2}.
\end{equation}
We extended this approximation $D(\vec{r}) = D_{mf}(\vec{r})+\Delta D(\vec{r})$ ($D_{mf}$ from Eq. \ref{eq:density}) through the fluctuation region to densities higher than $D_c$, until it smoothly matched up to the Thomas-Fermi approximation.

The method detailed above allows us to self-consistently calculate the density over any trapped sample. Since we measure total atom number in both of our planes, we find a value of $\mu^{global}$ which gives us (at the measured temperature) a calculated integrated density matching half of the total measured atom number. The peak phase space density shown in the figures in the main text is the calculated value of $D(0)$.

\bibliography{bibliography}

\begin{thebibliography}{10}

\bibitem{Thouless1973}
J~M Kosterlitz and D~J Thouless.
\newblock Ordering, metastability and phase transitions in two-dimensional
  systems.
\newblock {\em Journal of Physics C: Solid State Physics}, 6(7):1181--1203,
  1973.

\bibitem{Berezinskii1970}
V.L. Berezinskii.
\newblock {\em Sov. Phys. JETP}, 32:493--500, 1970.

\bibitem{Phillips2003}
Philip Phillips and Denis Dalidovich.
\newblock {The Elusive Bose Metal}.
\newblock {\em Science}, 302(5643):243--247, 2003.

\bibitem{Girvin1990}
Matthew P.~A. Fisher, G.~Grinstein, and S.~M. Girvin.
\newblock Presence of quantum diffusion in two dimensions: {U}niversal
  resistance at the superconductor-insulator transition.
\newblock {\em Phys. Rev. Lett.}, 64(5):587--590, Jan 1990.

\bibitem{fisher1990}
Matthew P.~A. Fisher.
\newblock Quantum phase transitions in disordered two-dimensional
  superconductors.
\newblock {\em Phys. Rev. Lett.}, 65(7):923--926, Aug 1990.

\bibitem{Markovic1998}
Allen~M. Goldman and Nina Markovic.
\newblock Superconductor-insulator transitions in the two-dimensional limit.
\newblock {\em Physics Today}, 51(11):39--44, 1998.

\bibitem{Bozovic2011}
A.~T. Bollinger, G.~Dubuis, J.~Yoon, D.~Pavuna, J.~Misewich, and I.~Bozovic.
\newblock Superconductor-insulator transition in
  {L}a$_{2-x}${S}r$_x${C}u{O}$_4$ at the pair quantum resistance.
\newblock {\em Nature}, 472:458--460, 2011.

\bibitem{Ong2010}
Lu~Li, Yayu Wang, Seiki Komiya, Shimpei Ono, Yoichi Ando, G.~D. Gu, and N.~P.
  Ong.
\newblock Diamagnetism and {C}ooper pairing above ${T}_{c}$ in cuprates.
\newblock {\em Phys. Rev. B}, 81(5):054510, Feb 2010.

\bibitem{Dalibard2006}
Z~Hadzibabic, P~Kruger, M~Cheneau, B~Battelier, and J~Dalibard.
\newblock {B}erezinskii-{K}osterlitz-{T}houless crossover in a trapped atomic
  gas.
\newblock {\em Nature}, 441(7097):1118--1121, JUN 28 2006.

\bibitem{Phillips2009}
P.~Clad\'{e}, C.~Ryu, A.~Ramanathan, K.~Helmerson, and W.~D. Phillips.
\newblock Observation of a 2{D} {B}ose gas: From thermal to quasicondensate to
  superfluid.
\newblock {\em Phys. Rev. Lett.}, 102(17):170401, 2009.

\bibitem{Inguscio2008}
Bose-{E}instein condensates in disordered potentials.
\newblock In E.~Arimondo, editor, {\em Advances In Atomic, Molecular, and
  Optical Physics}, volume~56, pages 119 -- 160. Academic Press, 2008.

\bibitem{Hulet2010}
D.~Dries, S.~E. Pollack, J.~M. Hitchcock, and R.~G. Hulet.
\newblock Dissipative transport of a {B}ose-{E}instein condensate.
\newblock {\em Phys. Rev. A}, 82(3):033603, Sep 2010.

\bibitem{Aspect2008}
J.~Billy, V.~Josse, Z.~Zuo, A.~Bernard, B.~Hambrecht, P.~Lugan, D.~Cl\`{e}ment,
  L.~Sanchez-Palencia, P.~Bouyer, and A.~Aspect.
\newblock Direct observation of {A}nderson localization of matter waves in a
  controlled disorder.
\newblock {\em Nature}, 453(7197):891--894, 2008.

\bibitem{Bouyer2010}
M.~Robert-de Saint-Vincent, J.-P. Brantut, B.~Allard, T.~Plisson, L.~Pezz\'e,
  L.~Sanchez-Palencia, A.~Aspect, T.~Bourdel, and P.~Bouyer.
\newblock Anisotropic 2d diffusive expansion of ultracold atoms in a disordered
  potential.
\newblock {\em Phys. Rev. Lett.}, 104(22):220602, Jun 2010.

\bibitem{DeMarco2009}
M.~White, M.~Pasienski, D.~McKay, S.~Q. Zhou, D.~Ceperley, and B.~DeMarco.
\newblock Strongly interacting {B}osons in a disordered optical lattice.
\newblock {\em Phys. Rev. Lett.}, 102(5):055301, Feb 2009.

\bibitem{Dalibard2009}
Z.~{Hadzibabic} and J.~{Dalibard}.
\newblock Two-dimensional bose fluids: An atomic physics perspective.
\newblock {\em ArXiv e-prints}, 0912.1490, dec 2009.

\bibitem{SmithBook}
C.J. Pethick and H.~Smith.
\newblock {\em Bose-Einstein Condensation in Dilute Gases}.
\newblock Cambridge University Press, 2008.

\bibitem{Bouyer2006}
D.~Cl\'{e}ment, A.~F. Var\'{o}n, J.~A. Retter, L.~Sanchez-Palencia, A.~Aspect,
  and P.~Bouyer.
\newblock Experimental study of the transport of coherent interacting
  matter-waves in a 1{D} random potential induced by laser speckle.
\newblock {\em New Journal of Physics}, 8(8):165, 2006.

\bibitem{Chin2011}
Chen-Lung Hung, Xibo Zhang, Nathan Gemelke, and Cheng Chin.
\newblock Observation of scale invariance and universality in two-dimensional
  {B}ose gases.
\newblock {\em Nature}, 470:236--239, 2011.

\bibitem{Cornell2010}
S.~Tung, G.~Lamporesi, D.~Lobser, L.~Xia, and E.~A. Cornell.
\newblock Observation of the presuperfluid regime in a two-dimensional {B}ose
  gas.
\newblock {\em Phys. Rev. Lett.}, 105:230408, 2010.

\bibitem{Svistunov2000}
Yu. Kagan, V.~A. Kashurnikov, A.~V. Krasavin, N.~V. Prokof'ev, and B.V.
  Svistunov.
\newblock Quasicondensation in a two-dimensional interacting {B}ose gas.
\newblock {\em Phys. Rev. A}, 61(4):043608, Mar 2000.

\bibitem{Blakie2009}
R.~N. Bisset, M.~J. Davis, T.~P. Simula, and P.~B. Blakie.
\newblock Quasicondensation and coherence in the quasi-two-dimensional trapped
  {B}ose gas.
\newblock {\em Phys. Rev. A}, 79(3):033626, Mar 2009.

\bibitem{Dalibard2011}
Tarik Yefsah, R\'emi Desbuquois, Lauriane Chomaz, Kenneth~J. G\"unter, and Jean
  Dalibard.
\newblock Exploring the thermodynamics of a two-dimensional {B}ose gas.
\newblock {\em Phys. Rev. Lett.}, 107:130401, Sep 2011.

\bibitem{Bourdel2011}
T.~{Plisson}, B.~{Allard}, M.~{Holzmann}, G.~{Salomon}, A.~{Aspect},
  P.~{Bouyer}, and T.~{Bourdel}.
\newblock {Coherence properties of a 2D trapped {B}ose gas around the
  superfluid transition}.
\newblock {\em ArXiv e-prints}, October 2011.

\bibitem{Svistunov2001}
Nikolay Prokof'ev, Oliver Ruebenacker, and Boris Svistunov.
\newblock Critical point of a weakly interacting two-dimensional {B}ose gas.
\newblock {\em Phys. Rev. Lett.}, 87(27):270402, Dec 2001.

\bibitem{Ketterle1997}
M.~R. Andrews, C.~G. Townsend, H.-J. Miesner, D.~S. Durfee, D.~M. Kurn, and
  W.~Ketterle.
\newblock Observation of interference between two {B}ose condensates.
\newblock {\em Science}, 275(5300):637--641, 1997.

\bibitem{Demler2006}
Anatoli Polkovnikov, Ehud Altman, and Eugene Demler.
\newblock {Interference between independent fluctuating condensates}.
\newblock {\em Proceedings of the National Academy of Sciences},
  103(16):6125--6129, 2006.

\bibitem{Gunn1990}
D~K~K Lee and J~M~F Gunn.
\newblock {B}osons in a random potential: condensation and screening in a dense
  limit.
\newblock {\em Journal of Physics: Condensed Matter}, 2(38):7753, 1990.

\bibitem{Hallock2006}
D.~R. Luhman and R.~B. Hallock.
\newblock Third sound on {C}a{F}$_{2}$ films of varying roughness.
\newblock {\em Phys. Rev. B}, 74(1):014510, Jul 2006.

\bibitem{Goldman1998}
N.~Markovi\ifmmode~\acute{c}\else \'{c}\fi{}, C.~Christiansen, and A.~M.
  Goldman.
\newblock Thickness-magnetic field phase diagram at the
  superconductor-insulator transition in 2{D}.
\newblock {\em Phys. Rev. Lett.}, 81(23):5217--5220, Dec 1998.

\bibitem{Garland1991}
D.~C. Harris, S.~T. Herbert, D.~Stroud, and J.~C. Garland.
\newblock Effect of random disorder on the critical behavior of {J}osephson
  junction arrays.
\newblock {\em Phys. Rev. Lett.}, 67:3606--3609, Dec 1991.

\bibitem{Essam1980}
J~W Essam.
\newblock Percolation theory.
\newblock {\em Reports on Progress in Physics}, 43(7):833, 1980.

\bibitem{Demarco2010}
M.~Pasienski, D.~McKay, M.~White, and B.~Demarco.
\newblock A disordered insulator in an optical lattice.
\newblock {\em Nature Physics}, 6(9):677--680, 2010.

\bibitem{Inguscio2010}
B.~Deissler, M.~Zaccanti, G.~Roati, C.~D'Errico, M.~Fattori, M.~Modugno,
  G.~Modugno, and M.~Inguscio.
\newblock Delocalization of a disordered bosonic system by repulsive
  interactions.
\newblock {\em Nature Physics}, 6(5):354--358, 2010.

\bibitem{Vinokur1994}
G.~Blatter, M.~V. Feigel'man, V.~B. Geshkenbein, A.~I. Larkin, and V.~M.
  Vinokur.
\newblock Vortices in high-temperature superconductors.
\newblock {\em Rev. Mod. Phys.}, 66(4):1125--1388, Oct 1994.

\bibitem{Svistunov2002}
Nikolay Prokof'ev and Boris Svistunov.
\newblock Two-dimensional weakly interacting {B}ose gas in the fluctuation
  region.
\newblock {\em Phys. Rev. A}, 66(4):043608, Oct 2002.

\bibitem{Krauth2010}
Markus Holzmann, Maguelonne Chevallier, and Werner Krauth.
\newblock Universal correlations and coherence in quasi-two-dimensional trapped
  {B}ose gases.
\newblock {\em Phys. Rev. A}, 81:043622, Apr 2010.

\end{thebibliography}

\end{document}